\documentclass[]{iopart}
\usepackage{graphicx}
\usepackage{natbib}
\usepackage{longtable}
\usepackage{iopams}
\usepackage{indentfirst}
\usepackage{graphicx}
\usepackage{rotating}
\usepackage{subfig}
\usepackage{url}
\usepackage{amssymb}

\begin{document}

\title{A List of Galaxies for Gravitational Wave Searches}

\author{Darren J. White$^1$, E. J. Daw$^1$ and V. S. Dhillon$^1$}
\address{$^1$Department of Physics and Astronomy, University of Sheffield, Sheffield, UK S3 7RH \\}
\date{Submitted 2010 Oct 12}


\label{firstpage}

\begin{abstract}
We present a list of galaxies within 100\,Mpc, which we call the Gravitational Wave Galaxy Catalogue (GWGC), that is currently being used in follow-up searches of electromagnetic counterparts from gravitational wave searches. Due to the time constraints of rapid follow-up, a locally available catalogue of reduced, homogenized data is required. To achieve this we used four existing catalogues: an updated version of the Tully Nearby Galaxy Catalog, the Catalog of Neighboring Galaxies, the V8k catalogue and HyperLEDA. The GWGC contains information on sky position, distance, blue magnitude, major and minor diameters, position angle, and galaxy type for 53,255 galaxies. Errors on these quantities are either taken directly from the literature or estimated based on our understanding of the uncertainties associated with the measurement method. By using the PGC numbering system developed for HyperLEDA, the catalogue has a reduced level of degeneracies compared to catalogues with a similar purpose and is easily updated. We also include 150 Milky Way globular clusters. Finally, we compare the GWGC to previously used catalogues, and find the GWGC to be more complete within 100\,Mpc due to our use of more up-to-date input catalogues and the fact that we have not made a blue luminosity cut.
\end{abstract}

\maketitle
\section{Introduction}

Gravitational waves (GWs) are ripples in space-time caused by accelerating masses. Predicted by Einstein's theory of General Relativity\cite{TGR}, they have been indirectly detected via observations of binary pulsars (see Weisberg \&\ Taylor 2004\nocite{pulsar}) but have so far eluded direct detection. The largest network of ground-based GW detectors, made up of the LIGO, Virgo and GEO interferometers, is now operating at sensitivities where GW signals could be directly detected (Abbott et al. 2009a; Acernese et al. 2008\nocite{abbott}\nocite{acernese}). The strongest GW signals are likely to come from inspirals, produced by merging neutron-star/black-hole binaries, and bursts, such as, for example, SNe, GRBs and magnetars\cite{candt2002}.

The detection of gravitational waves requires the identification of weak signals against strong noise backgrounds such as anthropogenic seismic noise and laser power fluctuations. Hence a promising approach for confirming the direct detection of GWs may come from associating a GW source with an electromagnetic (EM) counterpart. The position on the sky of a GW source can be estimated from the GW data. Depending on the source type, strength and position in the sky, the median positional accuracy is around 36 square degrees, but for strong events positional accuracies of down to $\sim 2^{\circ}$ are obtainable\cite{fairhurst09}. In one such scenario, a GW experiment detects a transient at a particular location and time in the sky, and then a telescope is immediately slewed to the same patch of sky, takes an EM image, and confirms the discovery by detecting a bursting source. As well as confirming the detection, the pre- (if available) and post-burst properties of the EM counterpart would enable detailed astrophysics of the GW source and its progenitor to be performed\cite{loocup}.

There are a couple of obvious problems with the above strategy. First, the best LIGO-Virgo position error circles are far larger than the widest fields of view obtainable from most large-aperture (i.e. 1m+) telescopes. Second, there are many EM transients in the sky -- how are we to know if a particular EM transient observed in the error circle of LIGO-Virgo is genuinely associated with the GW transient source.  Kulkarni \&\ Kasliwal (2009) find that foreground ``fog'' (asteroids, M-dwarf flares, dwarf novae) and background ``haze'' (distant, unrelated SNe) result in a significant chance of detecting an unrelated EM transient in a typical LIGO-Virgo error circle. Fortunately, the expected sensitivity of the LIGO \&\ Virgo interferometers in 2010 is a blessing in disguise here: it places a $\sim40$\,Mpc horizon on the majority of GW sources\cite{abbott}. This means that we can restrict EM follow-up to only the galaxies present within the LIGO-Virgo error circle, as plausible GW transient sources (e.g. GRBs) are far more likely to be extragalactic than Galactic in origin. This reduces the foreground fog and background haze by three orders of magnitude (Kulkarni \&\ Kasliwal 2009).

Efforts to achieve this have already been attempted by \cite{cbcg}, with the publication of the Compact Binary Coalescence Galaxies (CBCG) catalogue, containing 38,757 galaxies. When published it was claimed to be the most complete catalogue of galaxies within $100$\,Mpc. Only galaxies with a blue luminosity of $L_B\geq10^{-3}L_{10}$ were included, where $L_{10}=10^{10}L_{B,\odot}$ and $L_{B,\odot}=2.16\times10^{33}$ ergs/s, which is calculated using $M_{B,\odot}=5.48$. This cut was made as it is argued that in the nearby universe the compact binary coalescence rate is expected to follow the star formation rate\cite{phinney91}, which is traced using blue light.

In this paper we describe the compilation of a new galaxy catalogue\footnote{http://www.darrenwhite.postgrad.shef.ac.uk/gwgc.html}, which we call the Gravitational Wave Galaxy Catalogue (GWGC), providing a more complete, up-to-date sample created from a variety of literature sources extending out to 100\,Mpc, which is as unbiased as possible to a particular type of source. The GWGC contains a total of 53,225 galaxies and 150 globular clusters. This catalogue is currently being used in the search for electromagnetic counterparts within the LIGO/Virgo collaboration\cite{loocup}, and several data analysis groups in the collaboration (e.g. Nuttall and Sutton, 2010\nocite{nuttall}). It is therefore important that a full description of the catalogue and its construction appears in a readily-accessible location in the refereed literature. In $\S$2 we detail the compilation of the catalogue and describe important parameters we include, as well as error calculations. In $\S$3 we discuss the completeness of the GWGC in comparison to the CBCG catalogue and the Sloan Digital Sky Survey\cite{SDSSDR1}. 

\begin{figure}
	\centering
	\includegraphics[width=0.8\textwidth]{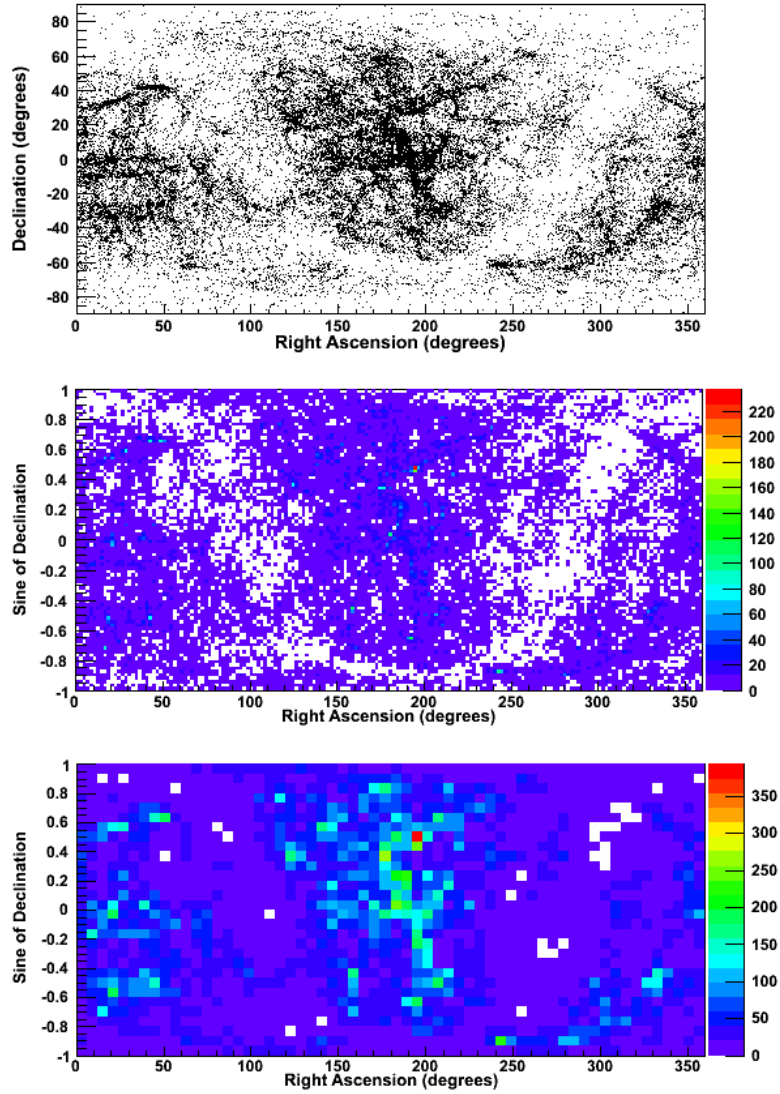}
	\caption{\small{Scatter plot showing distribution of galaxies in GWGC on the sky (top), and the distribution in $\sim$ 2 degree bins (middle) and 6 degree bins (bottom). The choice of bin size represents the best and median localisation of error circles on the sky obtainable with the LIGO/Virgo network of detectors\cite{fairhurst09}. This highlights the importance of using an available list of galaxies, as we must choose the best fields within a LIGO/Virgo error circle to observe.}}
	\label{fig:scatter}
\end{figure}
\section{Catalogue Compilation}

In order to improve on currently available catalogues, a larger, more up-to-date sample of galaxies is required. We also aim to improve the simplicity of incremental updates to the catalogue and minimise the risk of degeneracy within the catalogue itself, compared to similar catalogues. We achieved this by using scripts to create the GWGC from a subset of 4 large catalogues, each of which has a unique Principal Galaxy Catalogue (PGC) number for each galaxy\cite{pgc} These catalogues are: an updated version of the Tully Nearby Galaxy Catalog, the Catalog of Neighboring Galaxies, the V8k catalogue and HyperLEDA. These are available freely online but a local, homogeneous list is essential for rapid follow up purposes with LIGO and Virgo. A single, local catalogue also ensures that all working groups within the LIGO/Virgo Collaboration are using the same sources for both real-time and offline analyses. 

The Tully Nearby Galaxy Catalog\cite{tully3k}, hereafter referred to as Tully3000, is comprised of galaxies with a recession velocity $V < 3000$\,km s$^{-1}$ ($\sim$42\,Mpc for $H_0=72$\,km s$^{-1}$ Mpc$^{-1}$). Most recent release can be found at the Extragalactic Distance Database\cite{edd}\footnote{http://edd.ifa.hawaii.edu/}. A subset of these galaxies with high quality distance measurements were used to investigate the Local Void \cite{localvoid}. Tully3000 also contains the results of the Hubble Space Telescope (HST) Key Project to measure the Hubble constant\cite{hst}. 

The Catalog of Neighboring Galaxies, by \cite{cng}, is compiled from the literature and contains galaxies with a distance of $D\lesssim10$\,Mpc or a radial velocity of $V<550$\,km s$^{-1}$ ($D\lesssim7.6$\,Mpc for $H_0=72$\,km s$^{-1}$ Mpc$^{-1}$), and contains the less luminous, dSph and dIr galaxies often missed by larger surveys. 

The V8k catalogue\cite{edd} is another catalogue compiled from the literature, extending out to radial velocities $V < 8000$\,km s$^{-1}$.  Drawing heavily on the ZCAT survey\footnote{http://www.cfa.harvard.edu/$\sim$huchra/zcat/}, as well as other literature sources, the V8k catalogue excludes results from directional surveys such as the Sloan Digital Sky Survey (SDSS), Six Degree Field and Two Degree Field surveys, to provide a catalogue that is as uniform as possible across the sky out to a recession velocity of $V < 8000$\,km s$^{-1}$. HyperLEDA is also used to provide supplemental data where possible (for example, all positions angles are taken from HyperLEDA). We also include a list of 150 known Milky Way globular clusters\cite{mwgc}. Some of these are listed in HyperLEDA, and the respective PGC numbers are included where possible.

In order to be of use in the analysis and follow-up of gravitational wave data certain parameters are required to be accurately determined. These are the galaxy distances, diameters and blue magnitudes, as described below. Table \ref{tab:columns} shows the columns in the GWGC.

\begin{table}\small
\begin{center}
\begin{tabular}{cp{0.22\textwidth}p{0.48\textwidth}}\hline
Column & Abbreviation & Description \\ \hline
1 & {\bf PGC}& Identifier from HYPERLEDA \\
2 & {\bf Name}& Common name \\
3 & {\bf RA} & Right ascension (decimal hours) \\
4 & {\bf Dec} & Declination (decimal degrees) \\
5 & {\bf Type} & Morphological type code \\
6 & {\bf App\_Mag} & Apparent blue magnitude \\
7 & {\bf Maj\_Diam} & Major diameter (arcmins) \\
8 & {\bf err\_Maj\_Diam} & Error in major diameter (arcmins) \\
9 & {\bf Min\_Diam} & Minor diameter (arcmins) \\
10 & {\bf err\_Min\_Diam} & Error in minor diameter (arcmins) \\
11 & {\bf b/a} & Ratio of minor to major diameters \\
12 & {\bf err\_b/a} & Error on ratio of minor to major diameters \\
13 & {\bf PA} & Position angle of galaxy
(degrees from north through east, all $<180^{\circ}$) \\
14 & {\bf Abs\_Mag} & Absolute blue magnitude \\
15 & {\bf Dist} & Distance (Mpc) \\
16 & {\bf err\_Dist} & error on Distance (Mpc) \\
17 & {\bf err\_App\_Mag} & error on Apparent blue magnitude \\
18 & {\bf err\_Abs\_Mag} & error on Absolute blue magnitude \\
\hline
\end{tabular}
\end{center}
\caption{\label{tab:columns} \small{Data columns in the GWGC nearby galaxy catalogue.}}
\end{table}

\subsection{Spatial distribution of galaxies} 

The distribution of galaxies on the sky is far from uniform, as shown in fig. \ref{fig:scatter}. In the centre, the dense region is looking towards the Virgo cluster, and is also the primary region of observation in the SDSS survey. The empty region which traces a sinusoidal shape on the sky is the ``Zone of Avoidance'', in which gas and dust in the plane of the Milky Way obstructs our view. Offset by approximately +70 degrees in RA from the zone of avoidance, and with a similar shape, is the super-galactic plane, a sheet of galaxies in which the Virgo cluster and our galaxy reside. This is most clearly visible towards the bottom right of fig. \ref{fig:scatter}. On the left, we can also see strips of dense galaxies, which are due to the SDSS survey.

\subsection{Distances} 

\begin{figure*}
	\centering
	\subfloat[][]{\label{fig:gull}\includegraphics[width=0.47\textwidth]{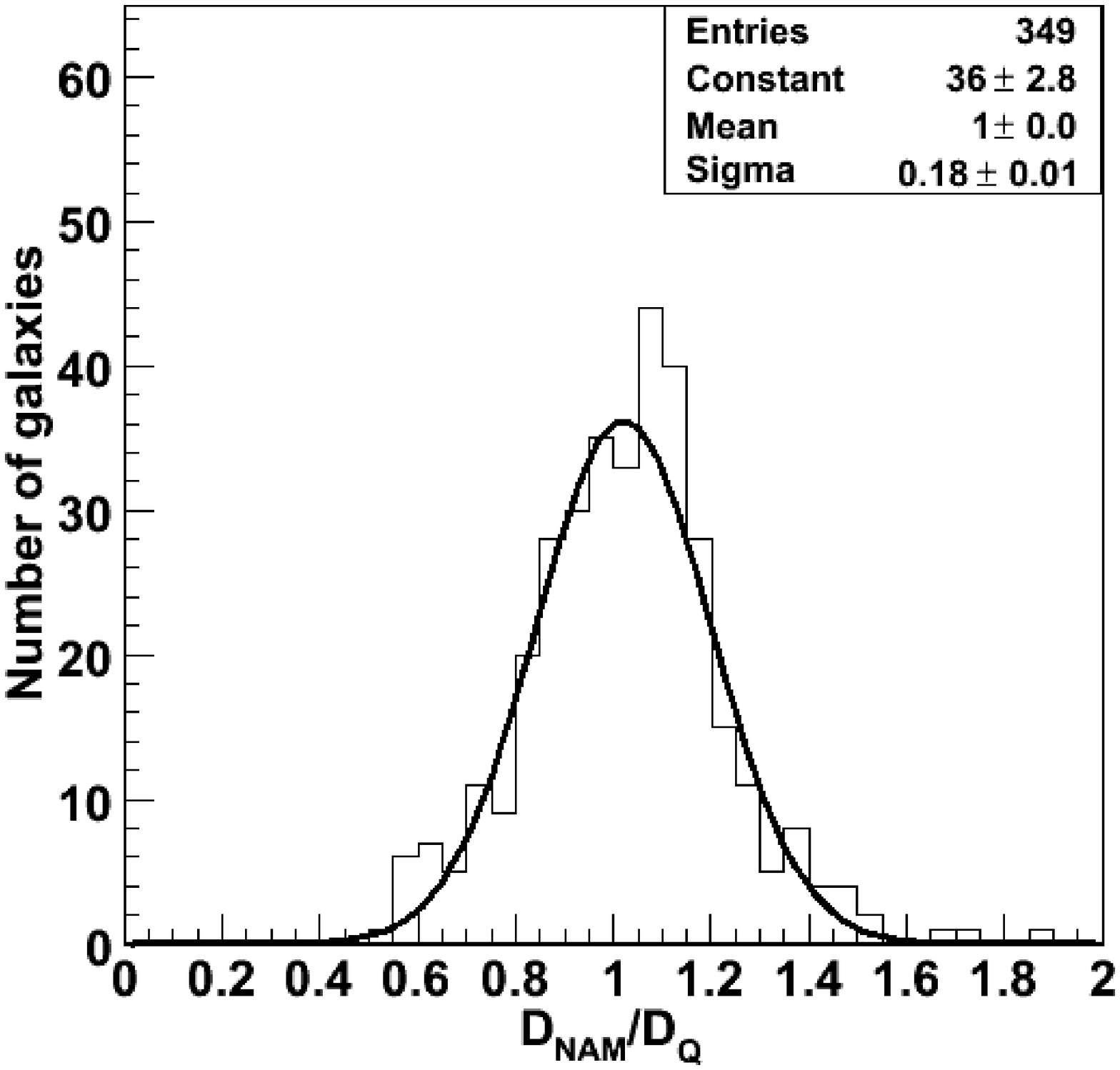}}
	\subfloat[][]{\label{fig:gull2}\includegraphics[width=0.47\textwidth]{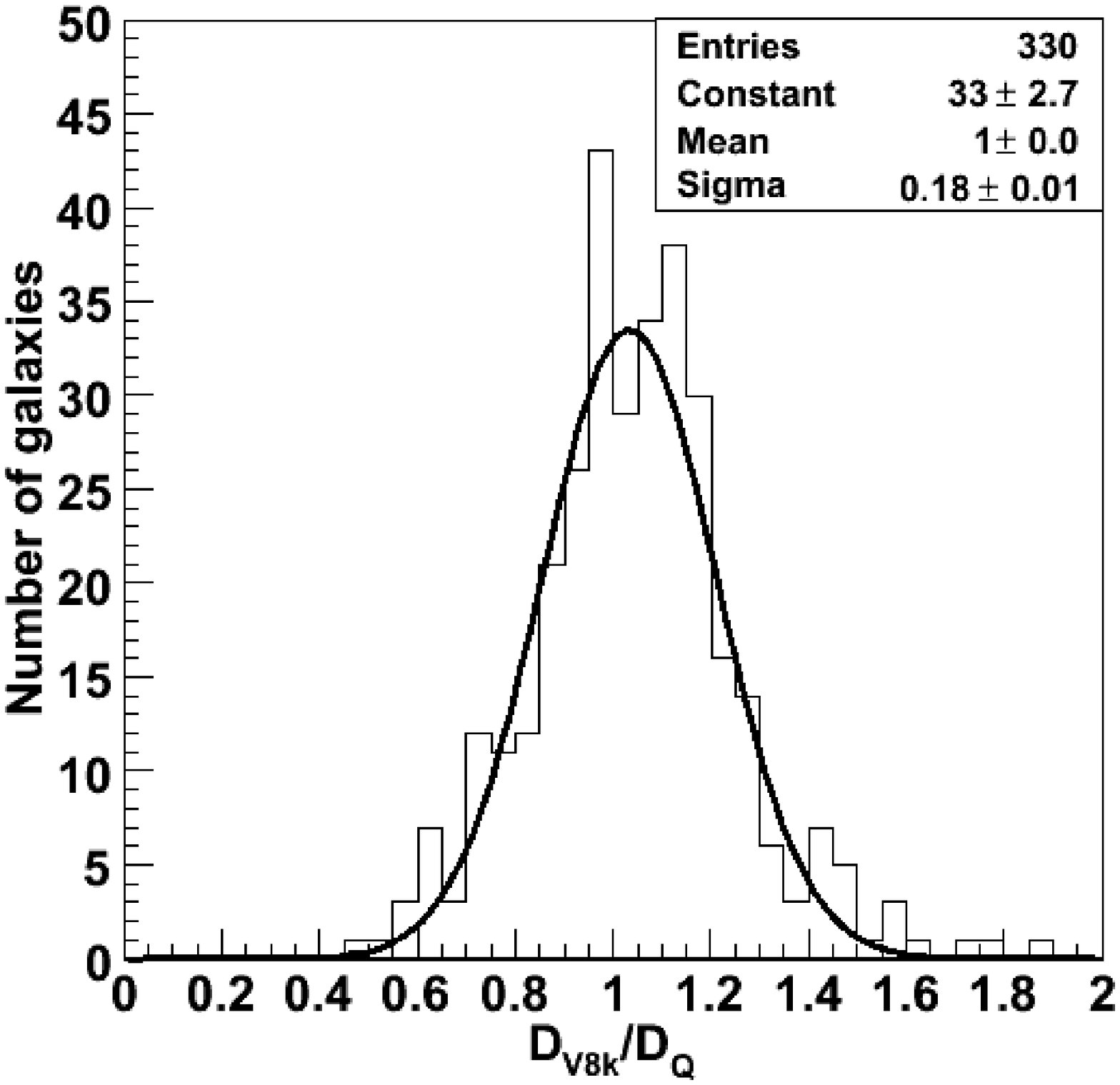}}\\
	\subfloat[][]{\label{fig:gull3}\includegraphics[width=0.47\textwidth]{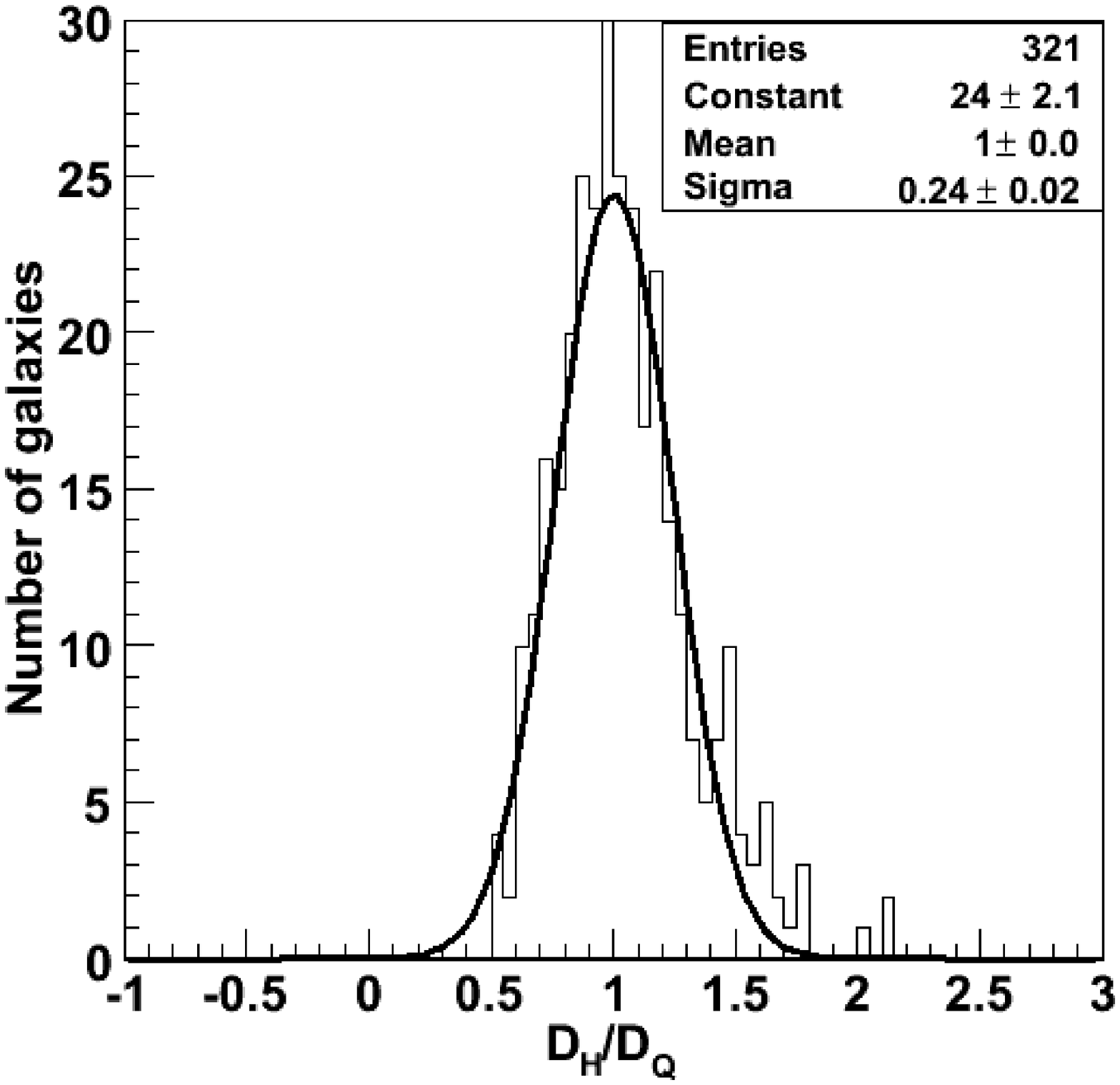}}
	\caption{\small{Comparison of distance measurements above 10\,Mpc for the same galaxies using different methods. Measurements in (a) are for the same galaxies appearing in the Tully3000 catalogue with both quality ($D_\mathrm{Q}$) and NAM ($D_\mathrm{NAM}$) distances. Those in (b) are for galaxies with $D_\mathrm{Q}$ distances from Tully3000 that also appear in the V8k catalogue ($D_\mathrm{V8k}$). (c) shows those galaxies with $D_\mathrm{Q}$ distances that appear in HyperLEDA with distances ($D_\mathrm{H}$) calculated using $v_\mathrm{vir}$.}}
	\label{fig:cats}
\end{figure*}

Accurate distances are important for electromagnetic follow-up observations as gravitational wave detectors have maximum distances at which expected sources are detectable. For example, the GWGC extends to 100\,Mpc but the maximum detectable distance for a 1.4\,M$_{\odot}$ binary neutron star inspiral is in the region of 30\,Mpc for current detectors\cite{LIGOdist}. An erroneous distance could cause a galaxy which is actually within 30\,Mpc to be missed. Similarly, it is possible that coincident detection of inspiral signals can be used to not only constrain position on the sky, but also a maximum and minimum distance to the source\cite{LIGOdist}, effectively giving us a region of space in which several possible host galaxies may lie. Inaccurate distance could again cause a galaxy to be missed. Ensuring that the galaxies in the GWGC have accurate distances is therefore vital. 

Each of the sub-catalogues in the GWGC contain distance measurements using a variety of methods, so we must estimate the accuracy of each method when not provided. The Tully3000 catalogue, which provides 3,496 galaxy distances to our resulting list, contains 3 groups of measurements: 

\begin{itemize}
	\item Distance measurements obtained using either the period-luminosity relation of Cepheid variable stars; the surface brightness fluctuation\cite{sbf}, where the amplitude of the luminosity fluctuation between pixels of a high signal-to-noise CCD image is inversely proportional to the distance; or the tip of the red giant branch\cite{trgb}, which uses the \textit{I}-band absolute magnitude of stars undergoing the helium flash stage of evolution, thought to be around -4.0. We refer to these collectively as `quality' distance estimates. The HST distance measurements in the Tully3000 catalogue have small ($<$10$\%$) errors. This is comparable to errors on the other quality measurements, which are estimated to be 10$\%$\cite{localvoid}.
	\item HI luminosity-line width distances measured using the Tully-Fisher (T-F) relation. Doppler broadening of the 21cm neutral hydrogen line is used to measure the rotational velocity of the galaxy, which is then used to estimate mass and, subsequently, luminosity. Combined with apparent magnitude, we can therefore estimate a distance. Distance measurements obtained using luminosity-line width observations are given an rms error of 20$\%$\cite{localvoid}.
	\item Distances converted from recession velocities using Hubble's law, corrected for infall towards the Virgo cluster using an evolved dynamical mass model of the local universe, the Numerical Action Model (NAM) by \cite{nam}, hereafter referred to as NAM distances, with errors calculated in the final paragraph of this section.
\end{itemize}

Distances in the Catalog of Neighboring Galaxies (an additional 120 galaxies) are measured using several methods: Cepheid variables, tip of the red giant branch, surface brightness fluctuation, T-F relation, brightest star luminosity, galaxy group membership and Hubble's law, in order of decreasing accuracy. Where more than one measurement is available for a galaxy the most accurate measurement is quoted. None of these are published with errors so we assign the same fractional errors to distances from the Catalog of Neighboring Galaxies using the same distnace measurement method as Tully3000. For sources where distance was measured using a method not in Tully3000 the distances are given fractional errors as calculated for galaxies taken from HyperLEDA, which we calculate in the final paragraph of this section.

Distances in V8k are primarily converted from redshift measurements using Hubble's law, after correcting for Virgo infall using the NAM model by \cite{nam}, as used for the Tully3000 catalogue. This catalogue, which provides 17,602 galaxies with distances within 100\,Mpc to the GWGC, also does not provide errors, so we must estimate them using the method used in the final paragraph of this section.

HyperLEDA only gives recession velocities corrected for infall towards the Virgo cluster ($v_{\mathrm{vir}}$, using a different model\footnote{see http://leda.univ-lyon1.fr/leda/param/vvir.html and references therein for full corrections}), without errors. We did not include any galaxies with $v_{\mathrm{vir}}\leq500$\,km s$^{-1}$ (7\,Mpc assuming $H_o=72$\,km s$^{-1}$ Mpc$^{-1}$) from HyperLEDA, as below this redshift-based distances are highly uncertain, and the use of Tully3000 and the Catalog of Neighboring Galaxies are thought to give us a high level of completeness in the local universe. From this catalogue, we included 32,007 galaxies with distances within 100\,Mpc.

Errors are strongly dependent on measurement method. However, only quality distances and T-F distances in Tully3000 and the Catalog of Neighboring Galaxies have error estimates based on distance measurement method. Therefore, in order to provide an error estimate for distances measured using other methods, we use galaxies which have multiple distance measurements. By plotting the ratio of two different distance estimates to the same galaxy and using a best fit Gaussian, we can determine the errors associated with methods with no published error estimates. Comparison between the quality measurements in Tully3000 and NAM distances give a best fit Gaussian with $\sigma=0.18$. Given that $\sigma=0.10$ for quality distance measurements, subtracting in quadrature gives fractional errors of 0.14 for NAM distances. Applying the same method to the V8k and HyperLEDA catalogues gives $\sigma=0.18$ and $\sigma=0.24$, respectively. This gives fractional errors of 0.15 for V8k distances and 0.22 for HyperLEDA distances.

\subsection{Blue luminosities}

Blue luminosity is a tracer of recent star formation, and in the nearby universe the distribution of binary neutron stars and black holes is expected to follow this star formation due to short merger timescales (Phinney 1991; Belczynski, Kalogera \& Bulik 2002\nocite{phinney91}\nocite{bco}). Therefore for investigations into the mergers of black hole and neutron star binary systems, blue luminosity is an important parameter to include. We have included both the apparent and absolute blue magnitudes where available. However, unlike the CBCG catalogue we have not applied a cut based on low blue luminosity, for two reasons. First, we do not wish to bias the GWGC towards any particular expected source. While blue light is a tracer of recent star formation, and therefore high mass compact binary coalescence and supernovae rates, there may be other sources of GW bursts. Second, as shown in \cite{defreitas}, the number of low blue-luminosity elliptical galaxies becomes significant in the Virgo cluster and beyond.

HyperLEDA has both absolute ($M_B$) and apparent ($m_B$) blue magnitudes corrected for Galactic extinction, internal extinction and K-correction with no errors. HyperLEDA also provides uncorrected apparent blue magnitudes with errors, which we use to apply the same fractional error to the corrected apparent blue magnitude. Tully3000 provides apparent blue magnitudes corrected for reddening, with an error of $\sigma (m_B)=0.3$ stated in \cite{cbcg} from a private communication with Tully, while the Catalog of Neighboring Galaxies only has uncorrected apparent blue magnitudes. Finally, the V8k catalogue provides absolute blue magnitudes corrected for reddening. We take Tully3000 magnitudes over HyperLEDA magnitudes, as these are fully corrected. If a galaxy has no corrected magnitude, we take the uncorrected magnitude if available. In total 49,364 ($\sim92.7\%$) of the galaxies in the GWGC have blue magnitude measurements.

For galaxies in the GWGC for which there is no error available for corrected apparent blue magnitude, we assign an error equal to the root of the mean of the square of the error estimates for the set of galaxies for which we do have published errors in HyperLEDA, which we find to be $\sigma (m_B)=0.43$. This is the mean of the square of the error estimates for those surveys that do publish an estimated magnitude of the errors. We have found that this is the best estimate we have of the error on the magnitude for the minority of galaxies for which we have no estimated error from the original survey.

\subsection{Angular Diameters and Position Angle}

A knowledge of the size, shape and orientation of the galaxies in the GWGC is essential in order to determine whether or not the galaxy fits within the field of view of a narrow field telescope. Methods such as drift scanning and mosaic imaging could increase the sizes of galaxies we image, but it is likely that electromagnetic counterparts to the expected sources of gravitational waves are going to be faint and very short lived. Therefore a rapid image of a whole galaxy is vital. In wide field follow up, the overlap of the galaxy with the LIGO/Virgo pointing is used as a weight to choose the best field to image. Therefore, in both narrow and wide field follow up planning, the size, shape and orientation of each galaxy is needed. In addition, this information can also be used with wide field image analysis to identify the location of the galaxy and constrain the transient search to those regions.

The Catalog of Neighboring Galaxies publishes major diameters ($a$) and the ratio of minor to major diameters ($b/a$), while the Tully3000 catalogue only publishes $b/a$ and the V8k catalogue only publishes $a$. HyperLEDA publishes $a$ and the ratio of major to minor diameters ($a/b$). HyperLEDA also provides position angle measurements. In the GWGC we include the major, minor and ratio of minor to major diameters, as well as the position angle where available. 

HyperLEDA is the only catalogue to provide errors on diameters and ratios, but using the same method used in \S2.3, we can estimate the errors on diameters for other catalogues. In HyperLEDA we find fractional errors $\sigma (a)/a=0.32$ and $\sigma (r)/r=0.12$ for diameter ratios. Globular clusters have diameters based on a variety of different measurements: the half mass radius, which is the distance in which half of the total mass of the cluster is contained; the core radius, which is the distance at which the surface brightness is 50\% of the centre of the cluster; and the tidal radius, which is the distance at which the globular cluster still has gravitational influence over the constituent stars. For the GWGC we use the tidal radius in order to include as much of the globular cluster as possible, with radius measurements available for 141 of the globular clusters. In total, the GWGC contains diameters for 47,179 ($\sim88.6\%$) galaxies and globular clusters.

\begin{figure}
	\centering
	\includegraphics[width=0.75\textwidth]{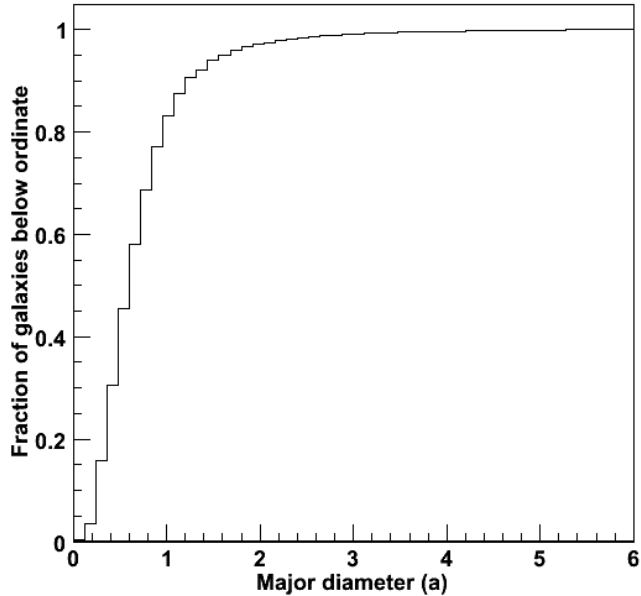}
	\caption{\small{Histogram of the major diameter of the galaxies in the GWGC}}
	\label{fig:majdiam}
\end{figure}

\section{Completeness} 

Observing faint galaxies in the local universe is a challenging task for any survey. Catalogue compilations will therefore suffer from incompleteness. The Catalog of Neighboring Galaxies, for example, is estimated to be $\sim80\%$ complete (Karachentsev et al. 2004\nocite{cng}), but only extends to a distance of $\sim10$\,Mpc. Analysis of luminosity functions can give and indication of the level of incompleteness in a catalogue. For the GWGC, we calculate the luminosity function as a function of distance, $N(M_{B},D)$, normalised to a spherical volume within radius $D$, in terms of absolute blue magnitudes using
\begin{equation} 
N(M_{B},D,\Delta M_{B})=\left(\frac{3}{4\pi D^{3}}\right) \sum_{j} l_{j} 
\label{eq:lumfunct}
\end{equation} 


\noindent where $l_{j}=1$ if $(M_{B}<M_{B,j}<M_{B}+\Delta M_{B})$ and $l_{j}=0$ otherwise. The index $j$ runs through all galaxies in the catalogue, where $D_{j}$ and $M_{B,j}$ are the distance and absolute magnitude of each galaxy. In order to investigate the completeness, we compare our luminosity function to the analytical Schechter galaxy luminosity function\cite{schech}, 
\begin{equation}
	\label{eq:schech1}
	\phi(L)dL=\phi^{\ast} \left(\frac{L}{L^{\ast}}\right)^{\alpha} \exp\left(\frac{-L}{L^{\ast}}\right) d\left(\frac{L}{L^{\ast}}\right)
\end{equation}

\noindent where $\phi(L)dL$ is the galaxy number density within the luminosity interval $L$ and $L+dL$, $L^{\ast}$ is the characteristic Schechter luminosity, the normalisation factor $\phi^{\ast}$ is the number density at the Schechter luminosity and $\alpha$ is the slope of the function at the faint end of the luminosity function. The last three of these must be determined empirically.
\begin{figure}
	\centering
	\includegraphics[width=0.75\textwidth]{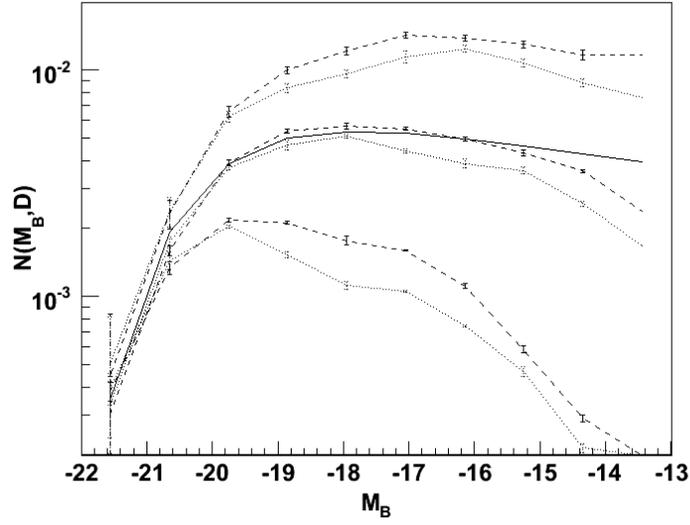}
	
	\caption{\small{Luminosity functions for both the GWGC (dashed) and the CBCG catalogue (dotted) at various distances: 20\,Mpc (top), 40\,Mpc (middle) and 100\,Mpc (bottom). The solid black line is the distance independant Schechter function from eq. \ref{eq:schech2}}}
	\label{fig:lumfunct}
\end{figure}

\subsection{Comparison to other results}

In terms of absolute blue magnitude eq. \ref{eq:schech1} becomes
\begin{equation}
	\centering
	\label{eq:schech2} 
	\begin{array}{ll}
	\phi(M_{B})dM_{B}&\!\!\!\!=0.92 \phi^{\ast} \exp\left[-10^{-0.4(M_{B}-M_{B}^{\ast})}\right]\\
	&\qquad\times [10^{-0.4(M_{B}-M_{B}^{\ast})}]^{\alpha+1} dM_{B}.\\
	\end{array}
\end{equation}

\begin{figure}
	\centering
	\includegraphics[width=0.75\textwidth]{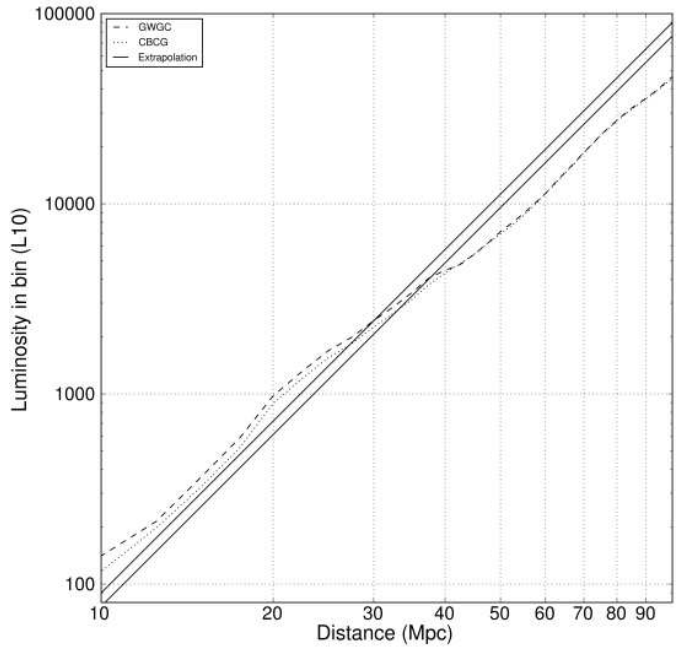}
	\caption{\small{Cumulative luminosity for the GWGC (dashed) and the CBCG catalogue (dotted) against distance, with extrapolation of blue luminosity density, with error (solid).}}
	\label{fig:cumlum}
\end{figure}

Using Table 2 in \cite{SDSSresults2}, and converting $g$-band to $B$-band using Table 2 in \cite{SDSSresults}, the Schechter parameters are $(M_{B}^{\ast},\phi^{\ast},\alpha)=(-20.3,0.0081,-0.9)$, from SDSS results extending to $z=0.1$, which are used to plot the Schechter function in fig. \ref{fig:lumfunct}. It can be seen that in comparison to the CBCG catalogue, the GWGC is more complete across all distances. By comparing both the shape and number density of our luminosity functions to the Schechter function, we can see evidence of where incompleteness occurs. Out to 20\,Mpc we see that our luminosity function has a similar shape, but higher number density, indicating that this is an overabundant region of space due to the Virgo cluster. At 40\,Mpc, our luminosity function follows the Schechter luminosity function closely until we reach galaxies fainter than approximately 15th magnitude. This is an indicator that we are incomplete at faint magnitudes. At 100\,Mpc, the difference between our luminosity function and the Schechter function is large, indicating a significant level of incompleteness, but we still offer improvement over the CBCG catalogue. 

We show the cumulative blue luminosity (in units of $L_{10}$) as a function of distance in fig. \ref{fig:cumlum} compared to the expected distribution of blue light if we assume a blue luminosity density of $(1.98\pm0.16)\times10^{-2}L_{10}$\,Mpc$^{-3}$ as calculated in \cite{cbcg} using SDSS results out to $z=0.1$. We also plot the CBCG catalogue for comparison. Using this method we find that the GWGC has completeness consistent with 100\% out to nearly 40\,Mpc, compared to just over 30\,Mpc for the CBCG catalogue. Comparing the cumulative blue luminosity of the GWGC to the extrapolation of the blue luminosity density at 100\,Mpc, we find that the GWGC is $\sim$60$\%$ complete. However, due to the non-uniform distribution of galaxies, this may not be completely representative of the true incompleteness of the catalogue. The Local Void\cite{localvoid} may, in part, explain some of the incompleteness beyond 40\,Mpc. However, due to the ``Zone of Avoidance'' we are certain to miss some galaxies. In future revisions we may attempt to overcome this problem with the inclusion of the 2MASS Redshift Survey (2MRS)\footnote{http://www.cfa.harvard.edu/$\sim$huchra/2mass/} or the 2MASS Galaxy Redshift Survey (2MASS XSCz)\footnote{http://web.ipac.caltech.edu/staff/jarrett/XSCz/index.html}, upon public release. These will help cover the ``Zone of Avoidance'' as well as extend the current catalogue out to greater distances, which is required for more sensitive GW detectors in the future. We must also stress that using blue luminosity distribution from a directional survey such as the SDSS may not provide the best estimate of completeness over the whole sky in the nearby universe, but it has been calculated in this paper to provide comparison to lists of galaxies made for a similar purpose.

\section{Conclusions}

In order to increase the likelihood of detecting electromagnetic counterparts to gravitational wave sources a complete catalogue of nearby galaxies is vital. Using a combination of local and extended galaxy catalogues from the literature, we have compiled a new catalogue reaching out to 100\,Mpc. For each galaxy we provide the most accurate distances and positions available, along with diameters, position angles and blue magnitudes where possible. We also provide errors on distances, diameters and magnitudes, either from the literature or estimated as described in the relevant sections. Comparing our galaxy catalogue to the expected distribution of blue light based on SDSS data shows that the catalogue is almost complete out to a distance of $\sim$40\,Mpc, but suffers from systematic incompleteness beyond this distance. This will only be truly solved with the inclusion of a deep, all-sky galaxy survey. The catalogue is also designed to be flexible, non degenerate and easily updated upon the release of new observations.

\section*{Acknowledgements}

We would like to thank Jonah Kanner, Erik Katsavounidis, Joel Fridriksson, Jeroen Homan, Marvin Rose, and the members of the Looc-Up LSC working group for many useful discussions. We would also like to thank B. Tully for discussion on nearby galaxy distributions. We acknowledge the use of the HyperLEDA database (http://leda.univ-lyon1.fr) and the Extragalactic Distance Database (http://edd.ifa.hawaii.edu/).

\label{lastpage}

\end{document}